\documentclass[10pt]{article}
\usepackage[letterpaper]{geometry}
\usepackage{hicss}
\usepackage{times}
\usepackage[none]{hyphenat}
\usepackage{url}
\usepackage{latexsym}
\usepackage{graphicx}
\usepackage{subcaption}
\usepackage{caption}
\graphicspath{{images/}}
\usepackage[
    style=ieee,
    maxnames=1,    
    minnames=1,    
    maxbibnames=2 
]{biblatex}
\usepackage{enumitem}

\usepackage{indentfirst}
\usepackage{xcolor}
\usepackage{amsmath}
\usepackage{graphicx}
\usepackage{float}
\usepackage{algorithmic}
\usepackage{array}
\usepackage{tabularx}
\usepackage{optidef} 
 
\usepackage{booktabs}
\usepackage{amssymb}
\usepackage{multirow}
\usepackage{makecell}
\usepackage{threeparttable}
\usepackage{indentfirst}
\usepackage{enumitem,kantlipsum}

\usepackage{xcolor}
\usepackage[ruled,vlined,linesnumbered]{algorithm2e}

\usepackage{booktabs}
\usepackage{multirow}
\usepackage{caption}
\usepackage{graphicx}
\usepackage{microtype}   
\SetKwComment{Comment}{// }{}

\usepackage{hyperref}
\usepackage{multicol}
\usepackage{fancyhdr}
\title{\textbf{Solving Three-phase AC Infeasibility Analysis to Near-zero Optimality Gap}}
\author{
    \begin{minipage}{0.45\textwidth}
        \centering
        Bikram Panthee \\
        University of Vermont \\
        \underline{bpanthee@uvm.edu}
    \end{minipage}%
    \hspace{0.1\textwidth} 
    \begin{minipage}{0.45\textwidth}
        \centering
        Amritanshu Pandey \\
        University of Vermont \\
        \underline{amritanshu.pandey@uvm.edu}
    \end{minipage}
    \vspace{-12mm}
}
\date{}
\vspace{-8mm}
\newcommand{\TPIA}{TPIA}
\newcommand{\BLP}{BL-TPIA}
\newcommand{\sbnb}{sBnB}
\newcommand{\sbt}{SBT}
\newcommand{\vdecomp}{variable decomposition}
\newcommand{\vfilt}{variable filtering}
\newcommand{\filx}{\mathbf{x}_f}
\newcommand{\ics}{infeasibility current sources}
\newcommand{\isc}{infeasibility source current}
\newcommand{\pblp}{\textbf{\emph{P}}_{\text{blp}}}
\newcommand{\nlp}{NLP}
\newcommand{\blp}{BLP}

\newcommand{\sblp}{S-BLP}

\newcommand{\ncases}{seven}
\newcommand{\rtime}{97}
\begin{document}
\maketitle
\thispagestyle{fancy} 
\begin{abstract}
\vspace{-1.0em}
Recent works have shown the use of equivalent circuit-based infeasibility analysis to identify weak locations in distribution power grids.
For three-phase power flow problems, when the power flow solver diverges, three-phase infeasibility analysis (TPIA) can converge and identify weak locations.
The original TPIA problem is non-convex, and local minima and saddle points are possible.
This can result in grid upgrades that are sub-optimal.
To address this issue, we reformulate the original non-convex nonlinear program (NLP) as an exact non-convex bilinear program (BLP).  
Subsequently, we apply the spatial branch-and-bound (\sbnb) algorithm to compute a solution with near-zero optimality gap. 
To improve \sbnb{} performance, we introduce a bound tightening algorithm with variable filtering and decomposition, which tightens bounds on bilinear variables.
We demonstrate that sequential bound tightening (SBT) significantly improves the efficiency and accuracy of Gurobi's \sbnb{} algorithm.  
Our results show that the proposed method can solve large-scale three-phase infeasibility analysis problems with $>5$k nodes, achieving an optimality gap of less than $10^{-4}$.
Furthermore, we demonstrate that by utilizing the developed presolve routine for bounding, we can reduce the runtime of \sbnb{} by up to \rtime{}\%.

\end{abstract}
\vspace{5pt}
\section{Introduction}

Distribution grids serve as the last-mile infrastructure for delivering electricity to households and commercial entities.
Due to rapid electrification and decarbonization, these grids are enabling increasingly higher electricity consumption, necessitating the need for upgrades and planning.
Utility planners run three-phase AC power flow analysis (TPPF) to identify and plan for grid upgrades.
However, as TPPF is a nonlinear problem, it is possible to have \textit{infeasible} three-phase AC networks.
In such scenarios, the TPPF solvers will diverge without indicating the source of failure.
Without access to any meaningful information, the grid planners cannot identify the weak or problematic locations in the network that cause infeasibility. Therefore, they cannot take meaningful corrective actions.

\cite{zamzam2016beyond, foster2022three, ali2024distributed} built three-phase infeasibility analysis (TPIA) framework to solve \textit{divergent} TPPF networks.
The approach solves an optimization rather than a simulation to solve \textit{divergent} TPPF problems while isolating weak locations.
However, due to its non-convex nature, it is hard to quantify the quality of the solution.
As a grid planner may rely on TPIA for \textit{optimally identifying} costly upgrades, it is necessary to develop methods that quantify the quality of TPIA solutions while avoiding local minima or saddle points.

Works that can certify near-zero optimality gaps for three-phase AC optimizations are limited.
Most existing approaches~\cite{gan2014convex, bernstein2017linear} rely on convex relaxations or linearizations to simplify the non-convex problem.
While these methods improve computational efficiency, they do not guarantee AC feasibility and are unsuitable for unbalanced distribution networks.
To the best of the authors' knowledge, no prior work has achieved near-zero optimality gap for large-scale unbalanced three-phase AC optimization problems.
In contrast, for balanced positive-sequence networks, prior works have shown to provide near-global solutions for a select set of networks.
For instance, \cite{lavaei2011zero} shows that semidefinite programming (SDP) relaxations can solve small-scale ACOPF instances (up to 300 buses) to global optimality.
Similarly, \cite{gopalakrishnan2012global, iranpour2025scalable} solves ACOPF by customizing the branch-and-bound method, which uses SDP relaxation and Quadratic Convex (QC) relaxations at each node to find near-global optimality conditions for small networks.

To solve the large-scale TPIA with a near-zero optimality gap, we utilize the spatial branch-and-bound (\sbnb{}) algorithm~\cite{smith1996global}.
The \sbnb{} algorithm is a well-known method for obtaining near-zero optimality gap solutions for non-convex bilinear problems.
Several studies~\cite{park2019convex, gopalakrishnan2012global, iranpour2025scalable} have applied this method to solve the ACOPF problem in transmission systems.
Although \sbnb{} can certify near-global solutions, it is difficult to scale to large-scale problems due to the exponential growth of the search space.
Recent work~\cite{das2024branch} has shown that tighter bounds on optimization variables significantly improve the efficiency of \sbnb{}.
Existing works~\cite{park2019convex, sang2024circuit} demonstrate the efficacy of optimization-based bound tightening techniques~\cite{nagarajan2016tightening} for transmission network problems. However, for large problems with many bilinear terms, the computation burden of tightening techniques can be onerous.

To address the current gaps and solve large-scale TPIA problems, we develop a presolving method (Section~\ref{subsec:presolve}) that improves the performance of \sbnb{} (in terms of runtime and node exploration) by generating tight variable bounds.
The main contributions of this work are as follows:
\begin{enumerate}[noitemsep, topsep=1.5pt]
    \item We present an exact bilinear reformulation of the \TPIA{} problem (with $L_1$ norm and $L_2$ norm), \BLP{}, enabling the application of spatial branch-and-bound (\sbnb{}) and sequential bound tightening (\sbt{}) methods.
    \item We introduce a variable filtering technique to minimize the number of variables to tighten during \sbt{}, and a decomposition technique to reduce the number of iterations per \sbt{} optimization.
   \item We validate our methodology on large-scale three-phase distribution networks ($>5$k nodes), achieving an optimality gap below $10^{-4}$, which can be considered globally optimal for all practical purposes.
\end{enumerate}
We evaluate our method on \ncases{} realistic but synthetic distribution feeders using the Gurobi \sbnb{} with a developed presolving routine.
The results (see section \ref{Experiments}) show that our presolving technique on the reformulated bilinear TPIA problem significantly reduces the number of explored nodes and solution time while providing near-zero optimality gaps.






\section{Three-phase Infeasibility Analysis (\TPIA{)}}

\begin{figure}[h]
    \centering
\includegraphics[width=1\linewidth]{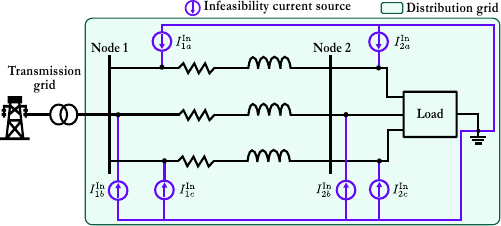}
    \caption{Illustration of a two-node three-phase distribution network with an infeasibility current source (purple) based on the $L_2$ norm formulation.}
    \label{fig:2_bus}
\end{figure}

\cite{foster2022three}, \cite{ali2024distributed} introduces three-phase infeasibility analysis (\TPIA{}) building on an earlier work for transmission networks \cite{jereminov2020evaluating}.
TPIA uses an equivalent circuit approach to solve the three-phase power flow problem (TPPF).
To also solve infeasible networks (networks with no power flow solution), it extends the simulation approach to an optimization approach.
TPIA modifies the network model by introducing a variable current source $I^{\text{In}}$ to all phases $p \in \phi$ for every node $n \in \mathcal{N}$ in the network. 
The variable current source can inject any current magnitude to balance the network equations.

Figure \ref{fig:2_bus} further illustrates the idea with infeasibility sources in the purple.
TPIA solves by minimizing the norm of these current sources subject to the network constraints.
If the TPIA optimization converges with an objective value of zero, the network is feasible, and the solution is consistent with the TPPF simulation.
Conversely, a nonzero objective value indicates that the network is infeasible and the non-zero \isc{}s localize areas of weakness.
Other variations to TPIA include variation in the type of infeasibility source from current sources to power sources ($P^{\text{In}}$ and $Q^{\text{In}}$) and impedance sources ($G^{\text{In}}$ and $B^{\text{In}}$).

\subsection{\TPIA{} Optimization Formulation}
\TPIA{} solves TPPF with no solution by adding \ics{} at each node.
The addition of \ics{} makes the problem underdetermined, allowing multiple possible solutions.
To obtain the TPPF solution if the network is feasible, and to find the smallest number of infeasibility source locations, when the network is infeasible, \TPIA{} solves an optimization problem (as in \eqref{NLP}) that minimizes the norm of \isc{}s while satisfying the three-phase AC network constraints and grid flow constraints.

\vspace{3mm}

\noindent \textbf{Objective}: We consider two objective functions for \TPIA{}: minimizing the \isc{}s with $L_2$ norm and $L_1$ norm, respectively:
\vspace{-1mm}
\begin{equation}\label{Objective_function_l2}
    f_{L_2}(I^{\text{In}} )  = \sum_{n \in \mathcal{N}} \hspace{4px} \sum_{p \in \Phi} \frac{1}{2} \|I_{np}^{r,\text{In}}\|_2^2 + \frac{1}{2}\|I_{np}^{i,\text{In}}\|_2^2 
\end{equation}
\begin{equation}\label{Objective_function_l1}
    f_{L_1}(I^{\text{In}} )  = \sum_{n \in \mathcal{N}} \hspace{4px} \sum_{p \in \Phi} |I_{np}^{r,\text{In}}| + |I_{np}^{i,\text{In}}| 
\end{equation}

\noindent where, $\mathcal{N} = \{1, 2, \dots, n, \dots\}$ represents the set of nodes, and $\Phi = \{a, b, c\}$ represents the set of phases.
The variables, 
$I^{r,\text{In}}$ and $ I^{i,\text{In}}$, represent the real and imaginary components of infeasibility current $I^{\text{In}}$,  respectively.

The $L_1$ norm formulation yields a sparse solution, which helps localize sources of grid weakness, while the $L_2$ norm spreads the infeasibility current injection throughout the network.

\vspace{2mm}

\noindent \textbf{Three-phase AC grid constraints:}
\TPIA{} includes three-phase AC power flow equations as equality constraints to ensure that solutions do not violate grid physics.
The three-phase unbalanced AC power flow equations, referred to as AC network constraints, are derived from Kirchhoff's Current Law (KCL), which ensures current balance at each node in the network.
 
In this work, we use the $IV$ formulation \cite{pandey2018robust},\cite{garcia2000three}. 
In this formulation, the nonlinearities are limited to injection nodes, and network constraints from lines and transformers are linear, making it more conducive to bilinear reformulations, as shown in Section \ref{sec:blp}.
Further, we represent real and reactive power ($P^{\text{load}}, Q^{\text{load}}$) inputs as functions conductances and susceptances ($G^{\text{load}}, B^{\text{load}}$) to help with reformulation in later sections. 
The compact form of network constraints are given by \eqref{kcl_real_constraint} - \eqref{gb_eequations}:
\begin{alignat}{1}
\forall n \in \mathcal{N}, \forall p \in \Phi:
\noindent \notag 
\end{alignat}
\underline{Real KCL constraints}\\
\begin{equation}\label{kcl_real_constraint}
    I_{np}^{r,\text{load}} + I_{np}^{r,\text{line}} - I_{np}^{r,\text{In}} = 0
\end{equation}
where,
\begin{subequations}\label{kcl_real_constraint__}
\begin{alignat}{1}
         & I_{np}^{r,\text{load}} - \left (G_{np}^{\text{load}} V_{np}^r - B_{np}^{\text{load}} V_{np}^i \right) = 0\label{Ir_load} \\
         & I_{np}^{r,\text{line}} - \left (\sum_{j \in \mathcal{N}} \hspace{4px}\sum_{\phi \in \Phi}G^{\text{line}}_{nj, p \phi} V_{nj, \phi}^r - B^{\text{line}}_{nj, p \phi}V_{nj,\phi}^i \right)= 0\label{Ir_line}
\end{alignat}
\end{subequations}
\underline{Imaginary KCL constraint}
\begin{equation}\label{kcl_imag_constraint}
    I_{np}^{i,\text{load}} + I_{np}^{i,\text{line}} - I_{np}^{i,\text{In}} = 0
\end{equation}
where,
\begin{subequations}\label{kcl_imag_constraint__}
\begin{alignat}{2}
         & I_{np}^{i,\text{load}} - \left (G_{np}^{\text{load}} V_{np}^i + B_{np}^{\text{load}} V_{np}^r \right) = 0 \label{Ii_load}\\ 
         & I_{np}^{i,\text{line}} - \left (\sum_{j \in \mathcal{N}} \hspace{4px}\sum_{\phi \in \Phi}G^{\text{line}}_{nj, p\phi } V_{nj, \phi}^i + B^{\text{line}}_{nj, p \phi}V_{nj,\phi}^r \right)= 0\label{Ii_line}
\end{alignat}
\end{subequations}
Equations \eqref{kcl_real_constraint} and \eqref{kcl_imag_constraint} represent Kirchhoff's Current Law (KCL) at each node $n$ and phase $p$.
$V_{np}^r$ and $V_{np}^i$ represent the real and imaginary voltage components at node $n$ and phase $p$.
Equations \eqref{Ir_load} and \eqref{Ii_load} represent the nonlinear real and imaginary load currents.
Equations \eqref{Ir_line} and \eqref{Ii_line} represent the linear real and imaginary line currents flowing away from node $n$, where $\phi$ denotes the mutual phases between lines.

The conductance $G_{np}^{\text{load}}$ and susceptance $B_{np}^{\text{load}}$ of the load at node $n$ and phase $p$ are expressed in terms of net active and reactive power consumption at that location and are defined as:
 \begin{subequations}\label{gb_eequations}
     \begin{alignat}{6}
         & G_{np}^{\text{load}} - \frac{P_{np}^{\text{load}}}{(V_{np}^r)^2 + (V_{np}^i)^2} = 0\label{G_load}\\
         & B_{np}^{\text{load}} + \frac{Q_{np}^{\text{load}}}{(V_{np}^r)^2 + (V_{np}^i)^2} = 0\label{B_load}
     \end{alignat}
 \end{subequations}

For notational simplicity, we omit three-phase transformers, shunt elements, switches, regulators, and fuses, as these linear components can be represented in a form analogous to the line equations.

\vspace{2mm}

\noindent \textbf{Grid Limit Constraints:}  
In \TPIA{}, we also enforce operational constraints on grid states.
Specifically, we ensure that voltage magnitudes remain within specified bounds and that line currents do not exceed their rated capacities.
\vspace{-2mm}
\begin{equation}\label{voltage_limit}
   \left(V^L_{np}\right)^2 \le \left(V_{np}^r\right)^2 + \left(V_{np}^i\right)^2 \le \left(V^U_{np}\right)^2
\end{equation}
\begin{equation}\label{current_limit}
     \left(I_{\psi}^{r}\right)^2 + \left(I_{\psi}^{i}\right)^2 \le \left(I^U_{\psi}\right)^2, \quad \forall \psi \in \mathcal{\Psi}
\end{equation}


\noindent In constraint \eqref{voltage_limit}, $V^L_{np}$ and $V^U_{np}$ denote the minimum and maximum voltage limits at node $n$ and phase $p$, respectively.  
In \eqref{current_limit}, $I^U_{\psi}$ is the thermal current rating for line $\psi$, with $\mathcal{\Psi}$ representing the set of lines. 
$I_{\psi}^{r}$ and $I_{\psi}^{i}$ denote the real and imaginary components of current on line $\psi$.  
\noindent The condensed form of \TPIA{} as a nonlinear program is shown in \eqref{NLP}:
\begin{subequations}\label{NLP}
    \begin{align}
        \textbf{\emph{P}}_{\text{nlp}}: \min \quad & f(I^{\text{In}}) \label{NLP_a} \\
        \text{s.t.} \quad & g(\mathbf{x}, I^{\text{In}} ) = 0 \label{NLP_b}\\
         & h(\mathbf{x}) \leq 0 \label{NLP_c}\\
        & \mathbf{x}^L \leq \mathbf{x} \leq \mathbf{x}^U \label{NLP_d}
    \end{align}
\end{subequations}
In \eqref{NLP}, the objective function $ f(I^{\text{In}})$ from \eqref{NLP_a} takes the form of either \eqref{Objective_function_l2} or \eqref{Objective_function_l1}.  
Constraint \eqref{NLP_b} represents the three-phase AC network equations, where $\mathbf{x}$ denotes the vector of all AC network TPPF state variables.
Constraint \eqref{NLP_c} enforces the grid limit constraints.
Constraint \eqref{NLP_d} bounds the state variables.
The AC network equations introduce nonlinearity and non-convexity into the optimization problem \eqref{NLP}, making the infeasibility analysis a non-convex Nonlinear Program (NLP).

The objective with the $L_2$ norm as in \eqref{Objective_function_l2} is differentiable, allowing us to use the gradient-based solvers to solve problem \eqref{NLP}.  
However, the objective with the $L_1$ norm in \eqref{Objective_function_l1} is non-differentiable due to the absolute value terms. 
To address this, \TPIA{} places the \ics{} as parallel current sources, 
with currents flowing in opposite directions, following the approach in~\cite{foster2022three}. Let $I_{np,+}^{r,\text{In}}$, $I_{np,-}^{r,\text{In}}$, $I_{np,+}^{i,\text{In}}$, and $I_{np,-}^{i,\text{In}}$ denote the non-negative real and imaginary infeasibility current sources, respectively nd defined as:
\begin{equation}\label{l1_ri}
    I_{np}^{r,\text{In}} = I_{np,+}^{r,\text{In}} - I_{np,-}^{r,\text{In}}, \quad
    I_{np}^{i,\text{In}} = I_{np,+}^{i,\text{In}} - I_{np,-}^{i,\text{In}}
\end{equation}

These parallel current sources are constrained to be non-negative:
\begin{equation}
    I_{np,+}^{r,\text{In}} , I_{np,-}^{r,\text{In}} , I_{np,+}^{i,\text{In}} , I_{np,-}^{i,\text{In}} \geq 0
\end{equation}
The objective function of \TPIA{} using the $L_1$ norm with parallel current sources, following the formulation in~\cite{foster2022three}, is given by:
\vspace{-2mm}
\begin{multline} \label{l1_new}
    \hat{f}_{L_1}(I^{\text{In}}) = \sum_{n \in \mathcal{N}} \hspace{4px} \sum_{p \in \Phi} 
    \bigg[ I_{np,+}^{r,\text{In}} + I_{np,-}^{r,\text{In}} + \\I_{np,+}^{i,\text{In}} + I_{np,-}^{i,\text{In}} \bigg]
\end{multline}
We replace the objective function \eqref{Objective_function_l1} with \eqref{l1_new} and modify the network constraint \eqref{NLP_b} by substituting the \ics{} with parallel current sources, as defined in \eqref{l1_ri}, to solve the \TPIA{} using the $L_1$ norm.



\section{TPIA Reformulation as a Bilinear Program}\label{sec:blp}


The original \TPIA{} problem is non-convex, and existing works solve it using local solvers, such as the primal-dual interior point method.
As such, it is hard to quantify the quality of these solutions as they may be local optima or saddle points.
Therefore, we study the efficacy of global solvers for these TPIA problems.
In particular, we examine the application of the spatial branch-and-bound (\sbnb{}) algorithm in Gurobi \cite{gurobi2025} for these optimizations.
As Gurobi's spatial branch-and-bound requires non-convexities to be limited to bilinearities, to apply outer approximation with McCormick Envelopes, we reformulate the original \TPIA{} into a bilinear program.
We further develop bound tightening and warm-starting routines to improve the efficacy of \sbnb{} in the context of the \TPIA{} problem.


\subsection{\BLP{} Mathematical Formulation} \label{subsec:blp_form}

In the original \TPIA{} problem, all equations—except \eqref{G_load} and \eqref{B_load}—are either linear, bilinear, or quadratic, and therefore do not require reformulation. Equations \eqref{G_load} and \eqref{B_load}, however, take the form of \eqref{nonlinear_y}:
\begin{equation}\label{nonlinear_y}
    s - \frac{a}{u^2 + v^2} = 0
\end{equation}
In this expression, $s$, $u$, and $v$ are variables, while $a$ is a known parameter.
To reformulate \eqref{nonlinear_y} as a bilinear constraint, we introduce a lifting variable $t$ defined by \eqref{w_eq}. This transformation lifts the problem from three dimensions to four dimensions.

\begin{equation}\label{w_eq}
    t - (u^2 + v^2) = 0 
\end{equation}

\noindent Using this lifting, we rewrite \eqref{nonlinear_y} in bilinear form as:
\begin{equation}\label{bilinear_y}
    st - a = 0
\end{equation}
Following the reformulation steps in \eqref{w_eq} and \eqref{bilinear_y}, we transform the original \TPIA{} formulation to the \BLP{} model by introducing a lifting variable $V^{\text{sq}}$, as defined in \eqref{vsq_equation}.
Using the lifted variable $V^{\text{sq}}$, we reformulate the nonlinear equations \eqref{G_load} and \eqref{B_load} into its bilinear form in \eqref{Vsq_kcl_G} and \eqref{Vsq_kcl_B}, respectively: 
\begin{equation}\label{vsq_equation}
    V^{\text{sq}}_{np} = (V_{np}^r)^2 + (V_{np}^i)^2
\end{equation}
\begin{equation}\label{Vsq_kcl_G}
    G_{np}^{\text{load}} \cdot V^{\text{sq}}_{np} - P_{np}^{\text{load}} = 0
\end{equation}
\begin{equation}\label{Vsq_kcl_B}
    B_{np}^{\text{load}} \cdot V^{\text{sq}}_{np} + Q_{np}^{\text{load}} = 0
\end{equation}
The resulting \BLP{} formulation has the following form:
\begin{subequations}\label{BLP}
    \begin{align}   \textbf{\emph{P}}_{\text{blp}}:\min \quad & f(I^{\text{In}}) \label{BLP_a} \\
        \text{s.t.} \quad & g_{\text{blp}}(\mathbf{x}, I^{\text{In}}, \mathbf{z}) = 0 \label{BLP_b}\\
        & h_{\text{blp}}(\mathbf{x}, \mathbf{z}) \le 0 \label{BLP_c}\\
        & \mathbf{x}^L \leq \mathbf{x} \leq \mathbf{x}^U \label{BLP_d}\\
        & z_k = x_ix_j \quad \forall k\in BL
    \end{align}
\end{subequations}
Here, $z_k$ denotes the $k^{\text{th}}$ bilinear term, where each $z_k = x_i x_j$ for $(i, j) \in BL$, and the vector $\mathbf{z}$ includes all such bilinear terms, i.e., $\mathbf{z} = \{z_k \mid k \in BL\}$.
The equality constraint \eqref{BLP_b} includes the bilinear AC network constraints.
The inequality constraint \eqref{BLP_c} represents the bilinear grid limit constraints.
The spatial branch-and-bound (\sbnb{}) algorithm~\cite{smith1996global} solves $\pblp$ in~\eqref{BLP} to global optimality by systematically partitioning the original feasible space (root node) through branching on continuous variables. It relaxes each resulting partitioned space (child node) using convex relaxations.
Effective performance of \sbnb{} relies on tight variable bounds from \eqref{BLP_d} (also used in McCormick relaxation in Section \ref{outer_mc}), which are essential both for branching decisions and relaxation accuracy.

When bounds are not provided, the solver attempts to infer them during presolve.
However, these inferred bounds are often loose, leading the algorithm to explore a large, less-informative feasible space.
This significantly degrades computational performance.
For large-scale bilinear programs, off-the-shelf solvers like Gurobi’s \sbnb{} become computationally impractical \cite{das2024branch}.

In contrast, providing tight \emph{a priori} bounds on bilinear terms, along with good initial conditions, significantly improves the efficiency of the \sbnb{} algorithm.
Therefore,  we introduce a presolving step that performs bound tightening (discussed in Section \ref{bound_tightening}) before solving \eqref{BLP}. 

\subsection{Presolving \BLP{}} \label{subsec:presolve}

We implement two strategies to improve the performance of \sbnb{} for \BLP{}: i) sequential bound tightening (\sbt{}) and ii) warm-starting.
\sbt{} can be computationally expensive as it requires solving two optimizations for every bilinear term.
We implement \vfilt{} and bound computation in Section \ref{filter} to reduce the number of \sbt{} optimizations.
We further enhance the tightening algorithm for the filtered variables by implementing \vdecomp{} in Section \ref{v_decompose}, which generalizes the initial bounds for filtered variables.
We also implement \sbt{} in parallel to improve performance.

\subsubsection{Variable filtering and bound computation} \label{filter}

The purpose of variable filtering is to identify a subset of variables, referred to as filtered variables and denoted by the $\mathbf{x}_f \subset \mathbf{x}$, for which bounds are tightened using \sbt{}. 
This subset is then used to compute bounds for the remaining bilinear terms, referred to as unfiltered variables and denoted by the $\mathbf{x}_{uf} \subset \mathbf{x}$. 
To implement \vfilt{}, we identify dependencies between bilinear variables in \BLP{}. 
Subsequently, we compute bounds for the dependent (unfiltered) variables analytically based on the filtered set.

For example, consider bilinear relationships in \eqref{w_eq} and \eqref{bilinear_y}, where variables $s$ and $t$ are functions of $u$ and $v$.
Given the best-known bounds on $u$ and $v$, we compute \textit{tighter} bounds for $t$ and $s$ as follows: 
\begin{subequations}\label{t_lb_ub}
    \begin{align}
        t^L &= \min(|u^L|, |u^U|)^2 + \min(|v^L|, |v^U|)^2 \\
        t^U &= \max(|u^L|, |u^U|)^2 + \max(|v^L|, |v^U|)^2
    \end{align}
\end{subequations}
\begin{subequations}\label{s_lb_ub}
    \begin{align}
        s^L &= \frac{a}{t^U} \\
        s^U &= \frac{a}{t^L}
    \end{align}
\end{subequations}
\noindent
Here, $(u^L, u^U)$ and $(v^L, v^U)$ denote the best-known lower and upper bounds of variables $u$ and $v$, respectively.
Similarly, $(t^L, t^U)$ and $(s^L, s^U)$ represent the computed lower and upper bounds of variables $t$ and $s$, respectively.
The use of the absolute function in \eqref{t_lb_ub} ensures that the computed lower and upper bounds of $t$ remain valid regardless of the sign of $u$ and $v$.

Without variable filtering, bound tightening must be performed on all four variables ($u, v, t, s$). 
With filtering, we only need to tighten the bounds for the independent variables $u$ and $v$, and can then compute the dependent bounds analytically through post-processing. 
This reduction significantly improves the performance of tightening in large-scale problems.

In the \BLP{} formulation, we assume $V^r$ and $V^i$ as independent variables, as we have good a priori knowledge of operating bounds for these terms.
Thus, the set of $V^r$ and $V^i$ forms the filtered variable set ($\mathbf{x}_f \subset \mathbf{x}$). 
Once the bounds of the filtered variables are known, we compute the bounds of the set of unfiltered variables, $V^{\text{sq}}, G^{\text{load}},$ and $B^{\text{load}}$, using equations \eqref{v_sq_lb}, \eqref{v_sq_ub}, \eqref{g_lb_ub}, and \eqref{b_lb_ub}, respectively.
\vspace{-1mm}
\begin{align}
    (V^{\text{sq}})^L &= 
    \min\left(|(V^r)^L|, |(V^r)^U|\right)^2 \nonumber \\
    &\quad + \min\left(|(V^i)^L|, |(V^i)^U|\right)^2 \label{v_sq_lb} \\
    (V^{\text{sq}})^U &= 
    \max\left(|(V^r)^L|, |(V^r)^U|\right)^2 \nonumber \\
    &\quad + \max\left(|(V^i)^L|, |(V^i)^U|\right)^2 \label{v_sq_ub}
\end{align}
\begin{equation} \label{g_lb_ub}
    (G^{\text{load}})^L = \frac{P^{\text{load}}}{(V^{\text{sq}})^U}, \quad
    (G^{\text{load}})^U = \frac{P^{\text{load}}}{(V^{\text{sq}})^L}
\end{equation}
\begin{equation}\label{b_lb_ub}
    (B^{\text{load}})^L = -\frac{Q^{\text{load}}}{(V^{\text{sq}})^L}, \quad
    (B^{\text{load}})^U = -\frac{Q^{\text{load}}}{(V^{\text{sq}})^U}
\end{equation}


\subsubsection{Variable Decomposition}\label{v_decompose}

We improve the performance of the \sbt{} algorithm by further normalizing the filtered variables $\mathbf{x}_f$ by separating it into its nominal value and a \textit{deviation} term:
\begin{equation}
    \mathbf{x}_f = \mathbf{x}_f^{\text{nom}} + \Delta \mathbf{x}_f
\end{equation}
In the \BLP{} formulation, the filtered variables ($\filx{}=\{V^r, V^i\}$) are functions of voltage magnitudes and phase angles.
For example, the real part of the voltage $V^r$ is defined as:
\begin{equation}
    V^r = V\cos{(\theta)}
\end{equation}
where $V$ is the nominal voltage magnitude, and $\theta$ denotes the phase angle.
For phases $a$, $b$, and $c$, $\theta$ takes values $0$, $-\frac{2\pi}{3}$, and $\frac{2\pi}{3}$ radians, respectively.

To estimate bounds on $V^r$, we need bounds on both $V$ and $\theta$.
To generalize across phases and nodes, we decompose the real and imaginary components of the voltage into a known nominal value (based on $V$ and $\theta$) and a voltage deviation factor ($\Delta V$), as shown in \eqref{Voltage_dev}:
\begin{subequations}\label{Voltage_dev}
\begin{align}
    V^r &= V\cos{(\theta)} + \Delta V^r, \\
    V^i &= V\sin{(\theta)} + \Delta V^i,
\end{align}
\end{subequations}
Here, $\Delta V^r$ and $\Delta V^i$ denote the deviation factors for the real and imaginary components of voltage, respectively. This decomposition enables the assignment of small, well-defined bounds to the set of deviation factors $\Delta \mathbf{x}_f$, which generalize across all phases. 
For instance, the real part of the current injection equation transforms as follows:
\begin{multline}\label{dev_form}
    I^{r,\text{load}} - \bigg[
        \underbrace{G^{\text{load}} \Delta V^r - B^{\text{load}} \Delta V^i}_{\text{Bilinear}} \\
        +\underbrace{G^{\text{load}} V^{\text{nom}} \cos(\theta^{\text{nom}}) - B^{\text{load}} V^{\text{nom}}\sin(\theta^{\text{nom}})}_{\text{Linear}}
    \bigg] = 0
\end{multline}


\subsubsection{BLP Outer Approximation with McCormick Envelopes} \label{outer_mc}

In sequential bound tightening, we iteratively formulate and solve an outer approximation of the BLP problem in \eqref{BLP}.
We use McCormick envelopes to formulate the outer approximation, and we use bounds based on \vfilt{} and \vdecomp{} techniques (see Section \ref{filter} and \ref{v_decompose}).
We show McCormick relaxation for a general bilinear term $st$ in \eqref{bilinear_y} with \eqref{mccormick}.
The McCormick relaxation \cite{mccormick1976computability} approximates the feasible region of this bilinear term using a convex outer envelope.
This approach replaces $st$ with a set of linear inequalities, two under-estimators and two over-estimators, which tightly bound its feasible region.
\begin{equation}\label{mccormick}
    \begin{aligned}
       s^U t + s t^U - s^U t^U &\le y \\
       s^L t + s t^L - s^L t^L &\le y \\
       s^U t - s t^L + s^U t^L &\ge y \\
       s^L t - s t^U + s^L t^U &\ge y
    \end{aligned}
\end{equation}
Following the relaxation in \eqref{mccormick}, we relax all bilinear terms in $\pblp$ (defined in \eqref{BLP}) using McCormick envelopes.
The bilinear terms in \eqref{Ir_load}, \eqref{Ii_load}, \eqref{vsq_equation}, \eqref{Vsq_kcl_G}, and \eqref{Vsq_kcl_B} involve voltage variables, which we decompose into nominal values and deviation terms.
This decomposition transforms these expressions into functions of the deviation factors $\Delta \mathbf{x}_f$.
We derive bounds on $ \mathbf{x}_f$ and $\mathbf{x}_{uf}$ using the best-known upper and lower bounds of $\Delta \mathbf{x}_f$ (see Section~\ref{filter}), and apply McCormick envelopes to approximate $\pblp$.

After applying McCormick relaxation to all bilinear terms, the infeasibility analysis problem with an $L_1$ norm objective reduces to a Quadratically Constrained Program (QCP). 
For the objective with $L_2$ norm, the problem becomes a Quadratically Constrained Quadratic Program (QCQP) due to the presence of quadratic terms in the objective.
Both formulations are convex.
The convex relaxation of the \BLP{} is given by:
\begin{subequations}\label{QP}
    \begin{align}
        \textbf{\emph{P}}_{\text{cvx}}:\quad 
        \min \quad & f(I^{\text{In}}) \label{QP_a} \\
        \text{s.t.} \quad 
        & g_{\text{aff}}(\mathbf{x}_{uf}, I^{\text{In}}, \Delta \mathbf{x}_{f}, \mathbf{z}) = 0 \label{QP_b} \\
        & h_{\text{cvx}}(\mathbf{x}_{uf}, \Delta \mathbf{x}_{f}, \mathbf{z}) \le 0 \label{QP_f} \\
        & h_{\text{mc}}(\mathbf{x}_{uf}, \Delta \mathbf{x}_{f}, \mathbf{z}) \le 0 \label{QP_c} \\
        & \mathbf{x}_{uf}^L \leq \mathbf{x}_{uf} \leq \mathbf{x}_{uf}^U \label{QP_d} \\
        & \Delta \mathbf{x}_{f}^L \leq \Delta \mathbf{x}_{f} \leq \Delta \mathbf{x}_{f}^U \label{QP_e}
    \end{align}
\end{subequations}
Here, constraint~\eqref{QP_b} represents the affine AC network equations.
Constraint~\eqref{QP_f} represents the convex grid limit equations.
Constraint~\eqref{QP_c} represents the McCormick envelopes on the bilinear terms.
Constraint~\eqref{QP_d} enforces bounds on the unfiltered variables, while constraint~\eqref{QP_e} bounds the deviation factors of the filtered variables.

\subsubsection{Bound Tightening}\label{bound_tightening}

The performance of the \sbnb{} algorithm depends on the size of the feasible region it explores. Best-known bounds on bilinear terms, which are often loose, enlarge the non-convex feasible space, necessitating extensive branching to identify incumbents and eventually the global solution. Although such bounds may simplify the problem formulation, they increase the solution time.
Tighter bounds reduce the search space and significantly enhance \sbnb{} efficiency.
Since variable bounds depend on problem-specific operating conditions, it is essential to compute bounds that are both tight and inclusive of the optimal solution region.

\begin{figure}[h]
    \centering
    \includegraphics[width=\linewidth]{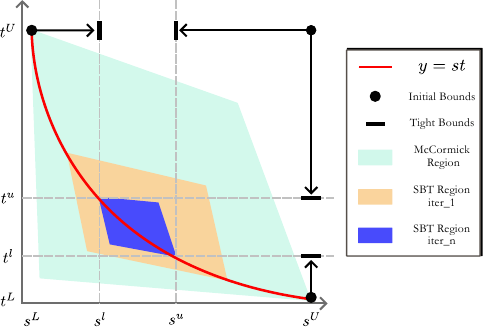}
    \caption{Conceptual illustration of \sbt{} applied to a bilinear term $y = st$. Starting from initial bounds $(s^L, s^U)$ and $(t^L, t^U)$, the \sbt{} find the tighter bounds $(s^l, s^u)$ and $(t^l, t^u)$ over $n$ iterations.}
    \label{fig:sbt_concept}
\end{figure}

To address this, we employ the Sequential Bound Tightening (\sbt{}) method proposed in~\cite{nagarajan2016tightening}, which builds on the iterative solving of the bound contraction method introduced in~\cite{CASTRO2015300}. 
Figure~\ref{fig:sbt_concept} illustrates how \sbt{} iteratively contracts loose bounds to yield tighter and more accurate bounds for bilinear variables.

More specifically, we tighten the bounds of the deviation factors $\Delta \mathbf{x}_{f}$ in constraint~\eqref{QP_e}.
By applying the \sbt{} method, we refine the initial bounds $(\Delta \mathbf{x}_{f}^L, \Delta \mathbf{x}_{f}^U)$ to tighter bounds $(\Delta \mathbf{x}_{f}^l, \Delta \mathbf{x}_{f}^u)$.
Using these improved bounds, we subsequently derive tighter bounds $(\mathbf{x}_{uf}^l, \mathbf{x}_{uf}^u)$ for the unfiltered variables $\mathbf{x}_{uf}$.

In Algorithm~\ref{Algm_1}, Lines 7--16 describe the \sbt{} procedure applied to the deviation factors $\Delta \mathbf{x}_{f}$. 
The \sbt{} method takes as input the initial bounds and a known feasible solution objective $f(I^{\text{In}^*})$ (Line 6) from $\textbf{\emph{P}}_{\text{nlp}}$. 
This feasible solution ensures that the tightened bounds do not eliminate feasible points with objective values lower bounding the $\textbf{\emph{P}}_{\text{nlp}}$ objective. 

\begin{figure}[h]
    \centering
    \includegraphics[width=1\linewidth]{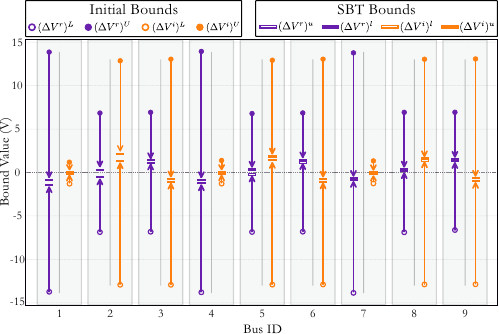}
    \caption{Reduction of initial bounds for real and imaginary voltage deviation terms using \sbt{} on the \texttt{GC-12.47-1} test case. On average, \sbt{} reduces the bounds of $\Delta V^r$ by 98\% and $\Delta V^i$ by 99\% within three iterations.
}
    \label{fig:sbt_reduction}
\end{figure}

For each element $\Delta x_{f,i}$, where $i = 1, \dots, k$, \sbt{} solves two optimization problems: one to minimize $\Delta x_{f,i}$ and another to maximize it, thereby identifying its tightest lower and upper bounds. 
These problems are solved subject to the constraints defined in \eqref{QP_b}--\eqref{QP_e}.
The bounds in constraints \eqref{QP_d} and \eqref{QP_e} are updated iteratively.
This process continues until the change in bounds of $\Delta \mathbf{x}_{f}$ is less than the specified tolerance $\epsilon$.
Moreover, \sbt{} allows parallel execution across all elements, enabling efficient bound tightening. 
Figure~\ref{fig:sbt_reduction} illustrates the significant reduction of bounds for the \texttt{GC-12.47-1} test case~\cite{schneider2008modern}.

\begin{algorithm}[t!]
\caption{{\bf Near-zero Optimality Gap TPIA} \label{Algm_1}}
\SetKwInOut{Input}{Input}
\SetKwInOut{Output}{Output}
\Input{3$\phi$ network, $I^{\text{In}}$ candidate locations, tolerance $\epsilon$, i/p bounds $({\mathbf{x}}^L, {\mathbf{x}}^U)$}
\Output{Non-zero $I^{\text{In}}$ magnitudes and locations, optimality gap}

\textbf{Reformulate} \TPIA{} as \BLP{}\\
\textbf{Filter} bilinear variables: assign filtered variables to $\mathbf{x}_f$ and unfiltered to $\mathbf{x}_{uf}$\\
\textbf{Decompose} $\mathbf{x}_f$ into nominal values $\mathbf{x}_f^{\text{nom}}$ and deviation terms $\Delta \mathbf{x}_f$\\
\textbf{Assign} initial bounds ($\Delta \mathbf{x}_f^L$, $\Delta \mathbf{x}_f^U$) for $\Delta \mathbf{x}_f$\\
\textbf{Compute} initial bounds ($\mathbf{x}_{uf}^L$, $ \mathbf{x}_{uf}^U$) for $\mathbf{x}_{uf}$\\
\textbf{Solve} $\textbf{\emph{P}}_{\text{nlp}}$ with local solver to obtain $f(I^{\text{In}^*})$\\

\BlankLine
\textbf{Run} Sequential Bound Tightening on $\Delta \mathbf{x}_f$:\\
\Input{$\Delta \mathbf{x}_f^L$, $\Delta \mathbf{x}_f^U$, $\epsilon$}
\textbf{Initialize:} $\Delta \mathbf{x}_f^l \leftarrow \Delta \mathbf{x}_f^L$, $\Delta \mathbf{x}_f^u \leftarrow \Delta \mathbf{x}_f^U$\\
\textbf{Set:}  $(\Delta \mathbf{x}_f)^l_{\text{prev}} \leftarrow 0$, $(\Delta \mathbf{x}_f)^u_{\text{prev}} \leftarrow 0$\\
\While{$\|\Delta \mathbf{x}_f^l - (\Delta \mathbf{x}_f)^l_{\text{prev}}\|_2 > \epsilon$ \textbf{and} $\|\Delta \mathbf{x}_f^u - (\Delta \mathbf{x}_f)^u_{\text{prev}}\|_2 > \epsilon$}{
    $(\Delta \mathbf{x}_f)^l_{\text{prev}} \leftarrow \Delta \mathbf{x}_f^l$, $(\Delta \mathbf{x}_f)^u_{\text{prev}} \leftarrow \Delta \mathbf{x}_f^u$\\
    \ForEach{$\Delta x_{f,i} \in \Delta \mathbf{x}_f$}{
        \textbf{Solve} in parallel:
        \begin{subequations}
        \begin{align}
            \Delta x_{f,i}^{l*} &:= \min (\Delta x_{f,i}) \\
            \Delta x_{f,i}^{u*} &:= \max (\Delta x_{f,i})
        \end{align}
        \text{subject to}
        \begin{align}
            &f(I^{\text{In}}) \leq f(I^{\text{In}^*}) \label{alg:sbt_obj_bound} \\
            &g_{\text{cvx}}(\mathbf{x}_{uf}, I^{\text{In}}, \Delta \mathbf{x}_f, \mathbf{z}) = 0 \label{alg:sbt_eq} \\
            & h_{\text{cvx}}(\mathbf{x}, \mathbf{z}) \le 0 \label{alg:sbt_g_limit}\\
            &h_{\text{mc}}(\mathbf{x}, \Delta \mathbf{x}_f, \mathbf{z}) \le 0 \label{alg:sbt_mcc} \\
            &\mathbf{x}_{uf}^l \le \mathbf{x}_{uf} \le \mathbf{x}_{uf}^u \label{alg:sbt_uf_bounds} \\
            &(\Delta \mathbf{x}_f)^l_{\text{prev}} \le \Delta \mathbf{x}_f \le (\Delta \mathbf{x}_f)^u_{\text{prev}} \label{alg:sbt_df_bounds}
        \end{align}
        \end{subequations}
    }
    $\Delta \mathbf{x}_f^l \leftarrow (\Delta \mathbf{x}_f)^{l*}, \quad \Delta \mathbf{x}_f^u \leftarrow (\Delta \mathbf{x}_f)^{u*}$\\
    \textbf{Update} ($\mathbf{x}^l_{uf}, \mathbf{x}^u_{uf}$)
}
\textbf{Return} tightened bounds: $\Delta \mathbf{x}_f^l , \Delta \mathbf{x}_f^u,\mathbf{x}^l_{uf}, \mathbf{x}^u_{uf}$\\
\textbf{Update} bounds on $\mathbf{x}$ with tightened bounds\\
\textbf{Solve} $\textbf{\emph{P}}_{\text{blp}}$ with \sbnb{} with tightened bounds and warm start with solution of $\textbf{\emph{P}}_{\text{nlp}}$  \\
\textbf{Report} Optimality gap and locations and magnitudes of non-zero  $I^{\text{In}}$
\end{algorithm}

Algorithm~\ref{Algm_1} describes the steps to improve the performance of the Gurobi \sbnb{} algorithm by applying presolve routines to the \BLP{} formulation before the Gurobi solve.
In algorithm~\ref{Algm_1}, line 1 reformulates the problem as a bilinear program.
Lines 2 and 3 apply variable filtering (from Section \ref{filter}) and decomposition (from Section \ref{v_decompose}).
Line 7 runs \sbt{} on the filtered variables (deviation term), $\Delta\mathbf{x}_f$, using McCormick Envelopes.
\eqref{alg:sbt_obj_bound} bounds the relaxation objective.
Line 15 tightens the bounds on unfiltered variables $\mathbf{x}_{uf}$ from SBT tightened variables $\Delta \mathbf{x}_f$.
With tightened bounds on the variables and the NLP warm-start solution, the algorithm runs the Gurobi \sbnb{} algorithm.
It then records the optimality gap and the corresponding locations of the \ics{}.
If the solver returns a solution with an optimality gap smaller than $\epsilon_{og}$, say $10^{-4}$, the algorithm certifies the solution as globally optimal for practical purposes.

\section{Experiments}\label{Experiments}

In this section, we describe the experimental setup for evaluating the effectiveness of the proposed presolving methodology for the \sbnb{} algorithm using the \BLP{} formulation.
We evaluate the scalability and solution quality of the proposed approach by testing it on \ncases{} synthetic distribution feeders from~\cite{schneider2008modern}, which vary in network size (quantified by the number of buses (\# Bus), as shown in Table~\ref{tab:results}).
The test cases, voltage bounds, and the weights are available at \url{https://github.com/pantheebikram/BL-TPIA#}.



\subsection{Experiment Setup}

We solve the bilinear problem $\textbf{\emph{P}}_{\text{blp}}$ using the Gurobi spatial branch and bound solver~\cite{gurobi2025}, version 12.0.2 and the IPOPT solver (version 3.14.17)~\cite{wachter2006implementation}.
All optimizations were run on the high-performance computing (HPC) cluster, which features an AMD EPYC 9654 96-core processor with 192 physical cores and 192 logical threads \cite{uvm_vacc}.
We solve the \BLP{} problem ($\pblp{}$) with three algorithms:
\begin{itemize}[noitemsep, topsep=1.5pt]
    \item \textbf{\nlp{}}: Primal-dual interior point with IPOPT
    \item \textbf{\blp{}}: Spatial branch and bound with Gurobi
    \item \textbf{\sblp{}}: Spatial branch and bound with pre-solve using sequential bound tightening and NLP warm start
\end{itemize}
We test two Gurobi configurations for \blp{} and \sblp{} algorithms.
We report the results for the configuration that has the best performance.
The first uses \textit{default} settings with only the \texttt{NonConvex} parameter set to 2 (specifying use of \sbnb{}).
The second applies\textit{ user-defined} settings: \texttt{NonConvex} = 2, \texttt{NumericFocus} = 3, \texttt{FeasibilityTol} = $10^{-5}$, \texttt{Presolve} = 1, \texttt{Heuristics} = 0.5, \texttt{MIPGap} = $10^{-4}$, \texttt{TimeLimit} = 36000 seconds, \texttt{Cuts} = 1, and \texttt{OBBT} = 3.
For IPOPT, we set the convergence tolerance to $10^{-7}$ and initialize the barrier parameter at $10^{-3}$, while retaining all other default settings.

\subsection{Evaluation}

To ensure \textit{realistic} evaluation across all algorithms, we assign a weight to each \ics{} before solving $\pblp{}$.
These weight will be user-defined, dependent on the cost of adding a corrective action at a certain location on the grid.
These weights are normalized such that their sum equals 1.
Table~\ref{tab:results} summarizes the results. 
We report the number of variables (\# Var), number of constraints (\# Constr), and the objective value (Objective) for all three algorithms and each test case.
For all algorithms except \nlp{}, we also report the best bound on the objective (Bestbd), the optimality gap percentage (Gap~(\%)), and the number of nodes (i.e., number of partitioned feasible space) explored by the \sbnb{} algorithm (\# Nodes).
Furthermore, for each algorithm and test case, we solve \BLP{} using both $L_1$ and $L_2$ norm objectives.

We evaluate the performance of the Gurobi \sbnb{} algorithm based on solution quality (measured by the objective value and optimality gap), the number of nodes explored, and the solution time.
Additionally, we compute the optimality gap for the \nlp{} algorithm using the following expression:
\begin{equation}
    \text{Gap (\%)} = \left| \frac{\text{Objective}_{\text{NLP}} - \text{Bestbd}}{\text{Objective}_{\text{NLP}}} \right| \times 100
    \label{eq:gap}
\end{equation}

For \nlp{}, the objective refers to the \textit{AC-feasible} solution obtained by IPOPT, and Bestbd corresponds to the best bound reported under the Gurobi \blp{} algorithm.

\begin{table*}[htbp]
\centering
\caption{Comparison of \sblp{} with \blp{} and \nlp{}}
\label{tab:results}
\resizebox{\textwidth}{!}{%
\begin{tabular}{@{}lllc*{6}{c}*{6}{c}@{}}
\toprule
\multirow{2}{*}{\textbf{Network}} & \multirow{2}{*}{\textbf{\#Bus}} & \multirow{2}{*}{\textbf{\#Constr}} & \multirow{2}{*}{\textbf{Algorithm}} &
\multicolumn{6}{c|}{\textbf{L1 Norm}} &
\multicolumn{6}{c}{\textbf{L2 Norm}} \\
\cmidrule(lr){5-10} \cmidrule(lr){11-16}
& & & &
\textbf{\#Var} & \textbf{Objective} & \textbf{Bestbd} & \textbf{\# Nodes} & \textbf{Time (s)} & \textbf{Gap} &
\textbf{\#Var} & \textbf{Objective} & \textbf{Bestbd} & \textbf{\# Nodes} & \textbf{Time (s)} & \textbf{Gap} \\
\midrule
\multirow{3}{*}{\textbf{GC-12.47-1}} & \multirow{3}{*}{93} & \multirow{3}{*}{567} & \nlp{} & \multirow{3}{*}{915} & 0.3359 & \dots & \dots & 0.04 & 0.0030\% & \multirow{3}{*}{735} & 1.5450 & \dots & \dots & 0.018 & 0.0092\% \\
 &  &  & \blp{} &  & 0.3359 & 0.3359 & 3 & 0.85 & 0.0021\% &  & 1.5450 & 1.5449 & 7 & 0.48 & 0.0092\% \\
 &  &  & \sblp{} &  & 0.3359 & 0.3359 & 1 & 0.2 & 0.0029\% &  & 1.5450 & 1.5449 & 1 & 0.08 & 0.0075\% \\ \cmidrule(lr){1-16}
\multirow{3}{*}{\textbf{R1-25.00-1}} & \multirow{3}{*}{830} & \multirow{3}{*}{5538} & \nlp{} & \multirow{3}{*}{8834} & 0.1768 & \dots & \dots & 0.459 & 0.0186\% & \multirow{3}{*}{7180} & 0.8798 & \dots & \dots & 0.306 & 0.0050\% \\
 &  &  & \blp{} &  & 0.1768 & 0.1768 & 6316 & 164.11 & 0.0030\% &  & 0.8798 & 0.8797 & 2183 & 92.99 & 0.0095\% \\
 &  &  & \sblp{} &  & 0.1768 & 0.1768 & 1 & 9.45 & 0.0078\% &  & 0.8798 & 0.8797 & 1 & 5.95 & 0.0037\% \\ \cmidrule(lr){1-16}
\multirow{3}{*}{\textbf{R4-25.00-1}} & \multirow{3}{*}{1274} & \multirow{3}{*}{6596} & NLP & \multirow{3}{*}{11668} & 0.0088 & \dots & \dots & 0.388 & 0.5370\% & \multirow{3}{*}{9126} & 0.0046 & \dots & \dots & 0.237 & 0.0416\% \\
 &  &  & BLP & & 0.0088 & 0.0088 & 5259 & 461.15 & 0.0021\% &  & 0.0046 & 0.0046 & 2850765 & 11763.46 & 0.0100\% \\
 &  &  & S-BLP &  & 0.0088 & 0.0088 & 736 & 151.35 & 0.0096\% &  & 0.0046 & 0.0046 & 35260 & 344.73 & 0.0100\% \\ \cmidrule(lr){1-16}
\multirow{3}{*}{\textbf{R4-12.47-2}} & \multirow{3}{*}{1680} & \multirow{3}{*}{8981} & NLP & \multirow{3}{*}{15677} & 0.1122 & \dots & \dots & 2.073 & 0.0571\% & \multirow{3}{*}{12323} & 0.4584 & \dots & \dots & 0.506 & 0.0099\% \\
 &  &  & BLP & & 0.1121 & 0.1121 & 1 & 174.85 & 0.0059\% &  & 0.4584 & 0.4584 & 21268 & 265.8 & 0.0095\% \\
 &  &  & S-BLP &  & 0.1121 & 0.1121 & 1 & 147.59 & 0.0077\% & & 0.4584 & 0.4584 & 7 & 15.14 & 0.0067\% \\ \cmidrule(lr){1-16}
\multirow{3}{*}{\textbf{R5-12.47-2}} & \multirow{3}{*}{1908} & \multirow{3}{*}{10921} & NLP & \multirow{3}{*}{18529} & 0.0845 & \dots & \dots & 5.643 & 0.1379\% & \multirow{3}{*}{14719} & 0.5548 & \dots & \dots & 1.943 & 0.1044\% \\
 &  &  & BLP &  & 0.0844 & 0.0844 & 610339 & \textcolor{red}{36000} & 0.0557\% &  & 0.5548 & 0.5542 & 1272629 & \textcolor{red}{36000} & 0.1040\% \\
 &  &  & S-BLP &  & 0.0844 & 0.0844 & 6560 & 996.48 & 0.0073\% &  & 0.5548 & 0.5545 & 1476857 & \textcolor{red}{36000} & 0.0541\% \\ \cmidrule(lr){1-16}
\multirow{3}{*}{\textbf{R2-12.47-3}} & \multirow{3}{*}{4943} & \multirow{3}{*}{25880} & NLP & \multirow{3}{*}{45628} & 0.0186 & \dots & \dots & 50.836 & 1.6214\% & \multirow{3}{*}{35748} & 0.0921 & \dots & \dots & 5.477 & 0.0220\% \\
 &  &  & BLP &  & 0.0183 & 0.0183 & 231900 & \textcolor{red}{36000} & 0.0261\% &  & 0.0921 & 0.0921 & 386729 & \textcolor{red}{36000} & 0.0160\% \\
 &  &  & S-BLP &  & 0.0184 & 0.0184 & 4866 & 6883.04 & 0.0000\% &  & 0.0921 & 0.0921 & 95818 & 9823.92 & 0.0097\% \\ \cmidrule(lr){1-16}
\multirow{3}{*}{\textbf{R1-12.47-1}} & \multirow{3}{*}{5194} & \multirow{3}{*}{26816} & NLP & \multirow{3}{*}{47568} & 0.1350 & \dots & \dots & 3.84 & 0.1334\% & \multirow{3}{*}{37186} & 2.8813 & \dots & \dots & 1.965 & 0.0028\% \\
 &  &  & BLP &  & 0.1348 & 0.1348 & 8688 & 352.05 & 0.0000\% &  & 2.8812 & 2.8812 & 14389 & 1544.33 & 0.0026\% \\
 &  &  & S-BLP &  & 0.1348 & 0.1348 & 1 & 45.45 & 0.0000\% &  & 2.8812 & 2.8812 & 11752 & 1260.52 & 0.0000\% \\
\bottomrule
\end{tabular}%
}
\end{table*}

\subsection{Analysis}

Table~\ref{tab:results} shows the outcomes for \ncases{} test cases solved using \nlp{}, \blp{}, and \sblp{} with both $L_1$ and $L_2$ norm formulations of \BLP{}.
We evaluate the computational efficiency of \sblp{} by analyzing the number of nodes explored by the \sbnb{} algorithm. 
With the exception of test case \texttt{R5-12.47-2} with the $L_2$ norm (which reached the specified time limit of 10 hr) \sblp{} consistently explores fewer nodes than \blp{}. 
The reduction is significant; for instance, in test case \texttt{R1-25.00-1}, where \sblp{} reduces the number of explored nodes by 99.98\%. 
Across all test cases, the average reduction in node count is 72.24\%.
This decrease in the number of explored nodes substantially reduces the solution time of \sbnb{}. 
In comparison to \blp{}, \sblp{} reduces the solution time by an average of 69.86\%. 
In some cases (see \texttt{R5-12.47-2} and \texttt{R2-12.47-3}), \blp{} fails to converge within 36,000 seconds (10 hours), whereas \sblp{} solves the problem within the time limit (showing the robustness of the algorithm); even though the current incumbent of the \blp{} seems to be the global solution; but the \blp{} algorithm has not explored the whole feasible space.

\begin{figure}[htbp]
    \centering
    \includegraphics[width=\linewidth]{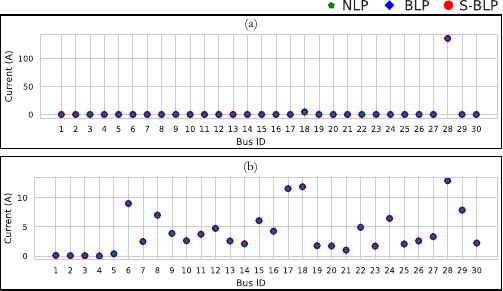}
    \caption{Comparison of \isc{} magnitudes from \nlp{}, \blp{}, and \sblp{} on the \texttt{GC-12.47-1} test case. (a) $L_1$ norm; (b) $L_2$ norm. \textbf{Note:} \texttt{Bus ID} denotes a 3-$\phi$ node; current magnitude is the norm of phase-wise currents.}
    \label{fig:ics_plots}
\end{figure}

Next, we assess the solution quality by analyzing the objective values and optimality gap percentages. 
We consider an optimality gap percentage of less than $10^{-4}$ to be a globally optimal solution for all practical purposes. 
For all test cases, except \texttt{R5-12.47-2} with the $L_2$ norm, \sblp{} achieves an optimality gap below the threshold of $10^{-4}$. Moreover, the objective values obtained from \sblp{} closely align with those from \blp{}, indicating that our presolving routine does not compromise solution quality.
For instance, we aggregate the infeasibility source magnitude at every node, and plot its value for all three algorithms in Figure~\ref{fig:ics_plots}.
We observe that \nlp{}, \blp{}, and \sblp{} solutions have the exact locations and the current magnitudes, demonstrating that the presolving routines developed in this paper preserve the physical interpretation of the solution.
The results also demonstrate the ability of the $L1$ norm to provide a sparse solution over the $L2$ norm objective (see Fig. \ref{fig:ics_plots}  top vs. bottom).

From the results in Table~\ref{tab:results}, we observe that the \nlp{} algorithm solves faster than both \blp{} and \sblp{}.
However, \nlp{} does not quantify solution quality and in rare case does seem to converge to a local solution (e.g., in case \texttt{R2-12.47-3}).

We evaluate the scalability of \sblp{} by varying the number of variables (735–47,568) and equality constraints (567–26,816). Despite the inherent non-scalability of the \sbnb{} algorithm, \sblp{} maintains performance by significantly reducing exploration space and solution time. As network size increases, more variables require filtering, which raises the computational cost of \sbt{}. Nevertheless, \sblp{} solves the large-scale \texttt{R1-12.47-1} case ($>$5k nodes, 47,568 variables) by exploring only a single \sbnb{} node.
A single run of \sbt{} for \texttt{R1-12.47-1} completes in an average of 11.18 seconds. As \sbt{} is parallelizable, its execution time depends primarily on the size of the available compute cluster. 
Notably, the computed bounds are reusable across multiple problem instances, further enhancing efficiency.
These findings highlight the effectiveness of our presolving strategy in enhancing the computational efficiency of the \sbnb{} algorithm while preserving solution accuracy.

\section{Conclusion}

We solved large-scale TPIA problems with near-zero optimality gap by reformulating the original \TPIA{} as a bilinear problem \BLP{} to apply the \sbnb{} algorithm.
We developed a presolving method to significantly improve the performance of the \sbnb{} algorithm. We draw the following key conclusions:

\begin{itemize}[noitemsep, topsep=1.5pt]
    \item The proposed \BLP{} formulation solved using \sbnb{} consistently achieves globally optimal solutions for all practical purposes, with optimality gaps below $10^{-4}$.
    \item The integration of sequential bound tightening (\sbt{}) significantly accelerates the \sbnb{} algorithm, reducing the total solution time by up to 97.23\%.
    \item The presolving routine also substantially reduces the number of nodes explored by the \sbnb{} algorithm, achieving a maximum reduction of 99.98\%.
\end{itemize}

\noindent Our work enables distribution grid planners to use \BLP{} results to take optimal corrective actions, such as adding capacitors, when three-phase AC networks are infeasible or divergent.

\section{Acknowledgement}
\vspace{-0.5em}
\noindent
This research was supported by the U.S. Department of Energy Vehicle Technologies Office (Grant No. DEEE0010640) and the Alfred P. Sloan Foundation (Grant No. 2024-22563).
We also thank Marko Jereminov for valuable discussions on the bilinear formulation.

\printbibliography
\end{document}